\documentstyle[PASJadd]{PASJ95}

\begin{document}
\setcounter{page}{0}
 
%
%-------------------------------------------------------------------
%

\title{Evolution of the Blue Luminosity-to-Baryon Mass Ratio
of Clusters of Galaxies}

\author{Kazuhiro {\sc Shimasaku} \\
{\it Department of Astronomy, University of Tokyo, 
7-3-1 Hongo, Bunkyo-ku, Tokyo 113 } \\
{\it Research Center for the Early Universe, University of Tokyo, 
7-3-1 Hongo, Bunkyo-ku, Tokyo 113} \\
{\it E-mail: shimasaku@astron.s.u-tokyo.ac.jp}}

%
%-------------------------------------------------------------------
%
 
\abst{
We derive the ratio of total blue luminosity to total baryon 
mass, $L_{\rm B}/M_b$, 
for massive ($M_{\rm gas}$ at the Abell radius is 
$\ge 1 \times 10^{13} h^{-2.5} \MO$) 
clusters of galaxies up to $z \simeq 1$ from the literature. 
Twenty-two clusters in our sample are at $z > 0.1$.
Assuming that the relative mix of hot gas and galaxies in clusters
does not change during cluster evolution, 
we use $L_{\rm B}/M_b$ to probe the star formation history 
of the galaxy population as a whole in clusters. 
We find that $L_{\rm B}/M_b$ of clusters increases with redshift 
from $L_{\rm B}/M_b=0.024 (L_{\rm B}/M)_\odot$ at $z = 0$ 
to $\simeq 0.06 (L_{\rm B}/M)_\odot$ at $z = 1$, 
indicating a factor of $2-3$ brightening
(we assume $H_0=70$ km s$^{-1}$ Mpc$^{-1}$).
This amount of brightening is almost identical to 
the brightening of the $M/L_{\rm B}$ ratio of early-type galaxies 
in clusters at $0.02 \le z \le 0.83$ reported by 
van Dokkum et al. (1998).
We compare the observed brightening of $L_{\rm B}/M_b$ 
with luminosity evolution models 
for the galaxy population as a whole, 
changing the $e$-folding time of star formation $\tau$ by 
$0.1 \le \tau \le 5$ Gyr  
and the formation redshift $z_{\rm F}$ by $2 \le z_{\rm F} < \infty$.
We find that 
$\tau = 0.1$ Gyr {\lq}single burst{\rq} models with $z_{\rm F} \ge 3$ 
and $\tau = 5$ Gyr {\lq}disk{\rq} models with arbitrary $z_{\rm F}$  
are consistent with the observed brightening,
while models with $\tau = 1-2$ Gyr tend to predict too steep 
brightening.
We also derive the ratio of blue luminosity density to 
baryon density for field galaxies, adopting $\Omega_b h^2 = 0.02$, 
and find that blue luminosity per unit baryon is 
similar in clusters and in fields up to $z \simeq 1$ 
within the observational uncertainties.
}
 
\kword{Galaxies: clusters of --- Galaxies: photometry --- 
Galaxies: X-rays }
 
\maketitle
\thispagestyle{headings}

%
%-------------------------------------------------------------------
%

\section{Introduction}

Clusters of galaxies are suitable objects for studying 
the evolution of galaxies in the dense environments.
Recent observations based on large telescopes 
including Hubble Space Telescope ($HST$) have been revealing 
the morphology-dependent 
evolution of galaxies in clusters up to $z \sim 1$.
The evolution of elliptical and S0 galaxies has been found 
to be reproduced well by the so-called single burst model 
(e.g., Schade, Barrientos, L\'{o}pez-Cruz 1997; 
Ellis et al. 1997; Kodama et al. 1998; van Dokkum et al. 1998).
Schade et al. (1996) found that spiral galaxies in clusters 
brighten by $\sim 1$ mag with redshift up to $z \simeq 0.5$, 
and that this brightening is similar to that of field 
spiral galaxies in the same redshift range 
(see, however, Vogt et al. 1997 and Lilly et al. 1998  
for the evolution of field spiral galaxies). 
Morphological studies based on $HST$ imaging suggest that 
a transition from spiral galaxies to S0 galaxies may have 
occurred in clusters since $z \sim 0.5$ (Dressler et al. 1997).
An attempt of measuring the star formation rate of individual 
galaxies has also started 
(e.g., Balogh et al. 1998; Poggianti et al. 1999).

In this paper, 
we study the global (or average) star formation history of 
the galaxy population as a whole in clusters of galaxies.
To do so, we derive the ratio of total blue luminosity ($L_{\rm B}$)
to total baryon mass ($M_b$) for clusters of galaxies 
up to $z \simeq 1$. 
The quantity $L_{\rm B}/M_b$, including its evolution, 
should reflect when (and what fraction of) baryons 
($=$ primordial gas) in clusters are converted into stars. 
Similar studies have been done for field galaxies.
The global luminosity density $l$ [$\LO {\rm Mpc}^{-3}$] 
in various wavelengths has been measured by many workers 
on the basis of observations of field galaxies 
(e.g., Lilly et al. 1996; Madau, Pozzetti, Dickinson 1998).
If the density parameter of baryons ($\Omega_b$) is given, 
one can compute from $l$ the mean luminosity per unit baryon mass 
in fields ($l/\rho_{\rm b}$).
We will derive $B$-band $l/\rho_{\rm b}$ 
in fields up to $z \simeq 1$ and compare it with $L_{\rm B}/M_b$ 
of clusters.

The structure of this paper is as follows.
In section 2, we present the data of nearby and distant clusters 
used to derive $L_{\rm B}/M_b$.
We compare $L_{\rm B}/M_b$ with predictions of simple luminosity 
evolution models in section 3. 
A comparison with $l/\rho_{\rm b}$ in fields 
is also given in section 3.
We summarize our conclusions in section 4.

We adopt $h=0.7$, $\Omega_0 = 0.2$, and 
$\lambda_0 = 0$ throughout this paper unless otherwise stated, 
where $h$ is the Hubble constant in units of 
$100$ km s$^{-1}$ Mpc$^{-1}$, 
$\Omega_0$ is the density parameter, and $\lambda_0$ is 
the cosmological constant.
Under this assumption, the present age of the universe is 
$11.8$ Gyr.
The value $h=0.7$ is taken from recent determinations 
of $H_0$ (e.g., Freedman 1999).
To adopt different values for $\Omega_0$ and $\lambda_0$ 
in the observationally reasonable ranges of 
$0.2 \le \Omega_0 \le 1$ and $0 \le \lambda_0 \le 0.8$ 
does not significantly change our results.

%
%---- section 2 ----------------------------------------------------
%

\section{Data}

We divide clusters of galaxies into nearby clusters ($z \le 0.1$) 
and distant clusters ($z > 0.1$).
We assume that evolutionary effects are negligible for $z \le 0.1$
clusters and regard their properties as those of 
the present-day clusters. 

We adopt the $B$ band to measure the luminosity of galaxies, 
and think that it is the best compromise.
Cluster luminosities have been measured mainly in optical bandpasses 
such as $B$, $V$, or $R$.
Among the optical bandpasses, $B$ is most sensitive to 
the luminosity evolution of galaxies. 
Though ultraviolet wavelengths such as $U$ are much better for 
measuring star formation, data in such wavelengths are very few.
As for field galaxies, 
there are a lot of measurements of luminosity density 
of field galaxies in the $B$ band, which enables us 
to compare $L_{\rm B}/M_b$ with $l_{\rm B}/\rho_b$.

%
%---- subsection 2.1 -----------------------------------------------
%

\subsection{Nearby Clusters}

We use the sample of nearby clusters given in 
Arnaud et al. (1992) to derive $L_{\rm B}/M_b$ 
of the present-day clusters.
Arnaud et al. (1992) compiled a sample of 27 clusters of galaxies, 
where total $V$-band luminosity ($L_{\rm V}$), 
morphological type mix of galaxies (E, S0, and S), 
and gas mass within a radius of $1.5 h^{-1}$ Mpc 
(the Abell radius) are given.
Morphological type mix is available for 18 clusters.

Arnaud et al. (1992) found in their clusters a strong dependence of
$L_{\rm V}/M_{\rm gas}$ on $L_{\rm V}$: 
$L_{\rm V}/M_{\rm gas} \propto L_{\rm V}^{-0.9}$.
If such a strong dependence holds in the whole mass range of 
clusters, it would very much complicate a comparison 
of $L_{\rm B}/M_b$ among clusters having different masses.
Thus, we first examine for what clusters such a strong dependence 
exists. 
Figure 1 plots $L_{\rm V}$ against $M_{\rm gas}$ 
for all the clusters in Arnaud et al. (1992).
The dependence found in Arnaud et al. (1992) is shown as 
the dashed line.
The solid line, on the other hand, is a regression line 
between $L_{\rm V}$ and $M_{\rm gas}$ 
for $M_{\rm gas} \ge 1 \times 10^{13} h^{-2.5} \MO$ clusters, 
$L_{\rm V} \propto M_{\rm gas}^{0.8}$.
This is close to a linear regression, i.e., 
a constant $L_{\rm V}/M_{\rm gas}$, indicated as the dotted line.
Thus, the strong dependence of $L_{\rm V}/M_{\rm gas}$ 
on $L_{\rm V}$ (or equivalently on $M_{\rm gas}$) 
found by Arnaud et al. (1992) 
is probably due to the inclusion of less massive clusters.

% Fig. 1

% Fig. 2

Figure 2 shows $L_{\rm V}/M_{\rm gas}$ as a function of 
the fraction of luminosity emitted from elliptical and S0 galaxies 
to the total luminosity for 18 clusters with type mix data.
It is found that $L_{\rm V}/M_{\rm gas}$ is constant 
for $M_{\rm gas} \ge 1 \times 10^{13} h^{-2.5} \MO$ clusters 
irrespective of type mix.
This implies that $L_{\rm V}/M_{\rm gas}$ is not sensitive to 
the change in the populations of galaxies for massive clusters.

Figures 1 and 2 demonstrate 
that the dependence of $L_{\rm V}/M_{\rm gas}$ 
on $L_{\rm V}$ (or on $M_{\rm gas}$) is much weak for massive clusters.
This is also supported by Renzini (1997) who found that 
rich clusters have a fairly constant $M_{\rm gas}$ to 
$B$-band luminosity ratio.
The reason why poor clusters have a relatively higher
$L_{\rm V}/M_{\rm gas}$ value is not clear, but a possible 
explanation is that a significant fraction of hot gas 
in poor clusters has escaped from the clusters 
during cluster evolution owing to their 
shallow gravitational potentials, resulting in 
a higher $L_{\rm V}/M_{\rm gas}$ value (e.g., Renzini 1997).

In any case, in what follows 
we assume that $L_{\rm B}/M_b$ is constant 
for $M_{\rm gas} \ge 1 \times 10^{13} h^{-2.5} \MO$ clusters 
and use them to derive the average $L_{\rm B}/M_b$ of nearby 
clusters.
We will see in the next subsection that all the distant clusters 
adopted in this paper have 
$M_{\rm gas} \ge 1 \times 10^{13} h^{-2.5} \MO$.
This promises a fair comparison of $L_{\rm B}/M_b$ 
between nearby and distant clusters.

For Arnaud et al.'s (1992) clusters which have morphological type mix, 
we compute total $B$-band luminosity ($L_{\rm B}$) from $L_{\rm V}$ 
using $B-V=0.96$ (E), $0.85$ (S0), and $0.68$ (S) 
(see Fukugita, Shimasaku, Ichikawa 1995).
For clusters without type mix data, 
we adopt $B-V = 0.85 \pm 0.2$ as the average color of galaxies.
We compute baryon mass $M_b$ from gas mass $M_{\rm gas}$ using: 
%
% equation 1
\begin{equation}
M_b = M_{\rm gas} + (M/L_{\rm B})_\star L_{\rm B},
\end{equation}
\noindent
where $(M/L_{\rm B})_\star$ is the mean mass-to-luminosity ratio of 
the stellar population in galaxies.
We neglect atomic and molecular gas in galaxies.
We adopt $(M/L_{\rm B})_\star = (6 \pm 3)h (M/L_{\rm B})_\odot$, 
which roughly covers the mass-to-luminosity ratio of 
elliptical galaxies (van der Marel 1991; Pizzella et al. 1997) 
and of spiral disks (Bahcall 1984; Broeils, Couteau 1997).
The errors in $L_{\rm B}/M_b$ contain  
(i) errors in $L_{\rm V}$ which are given in Arnaud et al. (1992),
(ii) errors in the mean $B-V$ (only for clusters without type mix 
data), 
and (iii) errors in $(M/L_{\rm B})_\star$ in equation (1).
Since most of the baryons in clusters are in form of hot gas, 
the error in $M_b$ due to (iii) is only about 5 \%.

% Fig. 3

Figure 3 presents $L_{\rm B}/M_b$ as a function of $M_{\rm gas}$.
The filled and open circles indicate Arnaud et al.'s (1992) 
clusters with and without type mix data, respectively.
As seen in figure 1, a clear trend is seen in figure 3 that 
clusters with $M_{\rm gas} \ltsim 1 \times 10^{13} h^{-2.5} \MO$ 
have systematically higher $L_{\rm B}/M_b$.
To derive $L_{\rm B}/M_b$ of nearby clusters, 
we not only remove clusters with 
$M_{\rm gas} < 1 \times 10^{13} h^{-2.5} \MO$ 
but also remove clusters without type mix data 
because the uncertainties in $L_{\rm B}$ of these clusters
are on the average larger than those for clusters having type mix 
(To include the clusters without type mix data hardly
changes the result, though).

Twelve out of the 21 clusters 
with $M_{\rm gas} \ge 1 \times 10^{13} h^{-2.5} \MO$ 
have type mix data, and their mean $L_{\rm B}/M_b$ for $h=0.7$ is 
%
% equation 2
\begin{equation}
L_{\rm B}/M_b = (0.024 \pm 0.004) (L_{\rm B}/M)_\odot,
\end{equation}
\noindent
which we regard as the representative value for the present-day 
clusters.
The contribution from elliptical and S0 galaxies 
to total $B$ luminosity is on the average $(69 \pm 13) \%$ 
for the 12 clusters, 
implying that these clusters are dominated by 
early-type galaxies (See figure 2).
Clusters which have $M_{\rm gas} < 1 \times 10^{13} h^{-2.5} \MO$
tend to be less dominated by early-type galaxies.
For example, the Virgo cluster, 
which has $M_{\rm gas} = 0.44 \times 10^{13} h^{-2.5} \MO$, 
has $L_{\rm B}({\rm E+S0})/L_{\rm B}({\rm tot}) = 45 \%$.

For the 12 clusters, we estimate the ratio of stellar mass 
to baryon mass to be 
$M_\star/M_b = 0.10 \pm 0.05$ ($h=0.7$), 
using $(M/L_{\rm B})_\star = (6 \pm 3)h (M/L_{\rm B})_\odot$.
This means that only $\sim 10$ \% of baryons have been 
used to form stars to date in rich clusters.

%
%---- subsection 2.2 -----------------------------------------------
%

\subsection{Distant Clusters}

Searching for total luminosity and gas mass data 
of distant clusters in the literature,
we take 22 clusters, among which 
thirteen are from the CNOC cluster sample 
(Carlberg et al. 1996; Lewis et al. 1999).
We do not apply any selection criterion to compile our sample.
Data of these clusters are given in table 1.
All the clusters have rest-frame luminosity 
(either $B$, $V$, or $r$ band) and gas mass measurements.
For a cluster whose rest-frame luminosity is in the $V$ or $r$ 
band, we use an observed or assumed color to convert the luminosity 
to the rest-frame $B$-band luminosity.

When computing $M_b$ from $M_{\rm gas}$ 
by $M_b = M_{\rm gas} + M_\star$, 
we use $M_\star/M_b = 0.10$ ($h=0.7$) which is 
the value for the nearby clusters.
Distant clusters may have lower $M_\star/M_b$ values than 
the nearby clusters, but the uncertainties in $L_{\rm B}/M_b$ due to 
this effect are at most $\simeq 10 \%$, 
which is negligible for our discussion.
Below are the references to the nine clusters and the CNOC clusters.

\vspace{5pt}
\noindent
(i) Abell 1413 ($z=0.14$) and Abell 1689 ($z=0.18$)

\noindent
These two clusters are taken from 
Cirimele, Nesci, Tr\`{e}vese's (1997) sample.
This sample consists of 12 clusters 
with $z < 0.2$ for which $M_{\rm gas}$ and $L_{\rm V}$ 
within a radius of $0.75 h^{-1}$ Mpc are given.
We calculate $L_{\rm B}$ assuming rest-frame $B-V = 0.85 \pm 0.2$.

\vspace{5pt}
\noindent
(ii) Abell 2218 ($z=0.18$)

\noindent
We adopt $M_{\rm gas}$ and $L_{\rm B}$ from Squires et al. (1996).

\vspace{5pt}
\noindent
(iii) Abell 2163 ($z=0.20$)

\noindent
$M_{\rm gas}$ and $L_{\rm V}$ are given in Squires et al. (1997).
$L_{\rm B}$ is computed assuming rest-frame $B-V = 0.8 \pm 0.2$.

\vspace{5pt}
\noindent
(iv) CL0500-24 ($z=0.32$)

\noindent
$M_{\rm gas}$ is taken from Schindler and Wambsganss (1997).
$L_{\rm V}$ is given in Infante et al. (1994), and 
$L_{\rm B}$ is computed using rest-frame $B-V = 0.85$ 
(The mean apparent color of this cluster, $V-I = 1.8$, 
reported by Infante et al. 1994
corresponds to $B-V=0.85$ in the rest frame).

\vspace{5pt}
\noindent
(v) CL0939+47 ($z=0.41$)

\noindent
Schindler et al. (1998) found two substructures in this cluster, 
implying that this cluster has not been virialized yet.
$M_{\rm gas}$ and $L_{\rm B}$ 
adopted here are the sum of the values for 
the two substructures given in Schindler et al. (1998).
Dressler et al. (1997) reported 
the fraction in number of elliptical and S0 galaxies to be $55 \%$.

\vspace{5pt}
\noindent
(vi) RXJ1347.5-1145 ($z=0.45$)

\noindent
$M_{\rm gas}$ is taken from Sahu et al. (1998) and $L_{\rm B}$ is 
computed from $M_{\rm tot}$ and $M_{\rm tot}/L_{\rm B}$ given in 
Fischer and Tyson (1997), where $M_{\rm tot}$ is the total mass 
of a cluster.

\vspace{5pt}
\noindent
(vii) CL0016+16 ($z=0.55$)

\noindent
$M_{\rm gas}$ is taken from Neumann and B\"ohringer (1997) 
and $L_{\rm B}$ is computed from the $r$-band total luminosity given 
in Carlberg et al. (1996) using rest-frame $B-r = 0.97$, 
which corresponds to the observed $g-r$ color of $1.455$ 
(Carlberg et al. 1996).
Dressler et al. (1997) reported 
the fraction in number of elliptical and S0 galaxies to be $73 \%$.

\vspace{5pt}
\noindent
(viii) AXJ2019+1127 ($z=1.01$)

\noindent
$M_{\rm gas}$ is taken from Hattori et al. (1997), and 
$L_{\rm B}$ is computed from $L_{\rm V}$ (Benitez et al. 1998) 
assuming rest-frame $B-V=0.7 \pm 0.3$, 
which roughly covers the expected colors of elliptical and 
spiral galaxies at $z=1$.

\vspace{5pt}
\noindent
{\it CNOC Clusters}

\noindent
Carlberg et al. (1996) give 
rest-frame $r$-band luminosity at the virial radius    
for 16 clusters at $0.17 < z < 0.55$.
Out of them, 14 clusters have gas mass measurements 
(figure 4 of Lewis et al. 1999).
Since the maximum radius at which gas mass is plotted, 
$r=600 h^{-1}$ kpc for all the clusters, 
is smaller than the virial radii, 
we derive for each cluster the $r$-band luminosity at $600 h^{-1}$ kpc 
from the value at the virial radius 
assuming that luminosity is proportional to radius.
The ratio of $600 h^{-1}$ kpc to the virial radii of 14 clusters is 
on the average $0.48$, implying that a large factor of 
conversion is necessary.
For each cluster, 
we then transform the rest-frame $r$-band luminosity 
into the rest-frame $B$-band luminosity 
on the basis of the observed $g-r$ color
given in figure 5 of Carlberg et al. (1996).
CL0016+16 is among the 14 clusters.
As seen in (vii), we have adopted for gas mass of this cluster 
the measurement given in Neumann and B\"ohringer (1997) 
because their value is at $r = 1.67 h^{-1}$ Mpc, 
which is very close to the virial radius 
where $r$-band luminosity is measured.
The number of the CNOC clusters adopted here is thus 13.

\vspace{5pt}
The radii used for measuring $L_{\rm B}/M_b$ differ among the clusters 
and most of them are smaller than the Abell radius
(See table 1). 
Unfortunately, it is not clear 
whether $L_{\rm B}/M_b$ measured at these small radii 
represent global values, i.e., values at the Abell radius, 
though there is a study that the $L_{\rm B}/M_b$ of the Coma cluster 
is nearly constant between a radius of $\simeq 0.4 h^{-1}$ Mpc 
and the Abell radius (Taguchi et al. 1999).
In this paper, we assume that the values of $L_{\rm B}/M_b$ derived 
here represent the global values of individual clusters 
(For CL0939+47, see, however, the next section).

% Fig. 4

Figure 4 plots $L_{\rm B}$ against $M_{\rm gas}$ for the 12 nearby 
clusters and the 22 distant clusters.
Both $L_{\rm B}$ and $M_{\rm gas}$ are values at 
$r = 1.5 h^{-1}$ Mpc, which are derived from raw values 
on the assumption that $L_{\rm B}(r)$ and $M_{\rm gas}(r)$ are 
proportional to radius $r$.
The thick solid line indicates the best fit of a linear law, 
$L_{\rm B} \propto M_{\rm gas}$, to the nearby clusters.
The thin solid line and the dotted line correspond to 
a similar fit to distant clusters at $0.1 < z \le 0.4$ and 
$0.4 < z \le 0.7$, respectively.
It is found that the average $B$ luminosity at a given gas mass 
increases with redshift.
This should reflect some evolution of $L_{\rm B}/M_{\rm gas}$.
Note that the range of gas mass is similar between 
the nearby and distant clusters: 
there is no distant cluster in our sample whose gas mass 
at $r = 1.5 h^{-1}$ Mpc 
is less than $1 \times 10^{13} h^{-2.5} \MO$.
Note also that no clear dependence of $L_{\rm B}/M_{\rm gas}$ 
on $M_{\rm gas}$ is seen either in the nearby cluster sample 
or in the distant cluster sample.

% Fig. 5

We mention here the effects of changing $\Omega_0$ and $\lambda_0$ 
on estimates of $L_{\rm B}/M_b$ for distant clusters.
Let $d_L(z)$ and $d_A(z)$ be the luminosity distance and 
the angular diameter distance to a cluster at $z$, respectively.
The ratio $L_{\rm B}/M_{\rm gas}$ is proportional to $d^{-0.5}_L(z)$, 
because of $L_{\rm B} \propto d^2_L(z)$,  
$M_{\rm gas} \propto d^{2.5}_A(z)$, and $d_L(z) = (1+z)^2 d_A(z)$.
Thus, $L_{\rm B}/M_{\rm gas}$ depends on $\Omega_0$ and $\lambda_0$ 
through $d^{-0.5}_L(z)$.
Since $M_b$ is dominated by $M_{\rm gas}$, the dependence of
$L_{\rm B}/M_b$ on $\Omega_0$ and $\lambda_0$ 
is very close to that of $L_{\rm B}/M_{\rm gas}$.
Figure 5 shows the ratio of $L_{\rm B}/M_{\rm gas}(\Omega_0,\lambda_0)$ 
to $L_{\rm B}/M_{\rm gas}(\Omega_0=0.2,\lambda_0=0)$ as a function 
of redshift for two sets of $(\Omega_0, \lambda_0)$.
Since $d_L(z)$ is a decreasing function of $\Omega_0$ and 
an increasing function of $\lambda_0$ up to at least $z=1$, 
we find from this figure that the change in $L_{\rm B}/M_{\rm gas}$ 
due to the change in $\Omega_0$ and $\lambda_0$ 
in the ranges of $0.2 \le \Omega_0 \le 1$ 
and $0 \le \lambda_0 \le 0.8$
is less than $\pm 10\%$ for $z<1$ clusters, 
which is negligible for our discussion below.

%
%---- section 3 ----------------------------------------------------
%

\section{Results and Discussion}

Figure 6 shows $L_{\rm B}/M_b$ of clusters as a function of redshift.
The filled circles present the distant clusters and 
the filled square indicates the average $L_{\rm B}/M_b$ 
of the nearby clusters.
Clusters without errors (but for CL0939+47 and CL0016+16) 
are the CNOC clusters.
Lines indicate model predictions, which will be 
discussed in the next subsection.
It is found that $L_{\rm B}/M_b$ increases with $z$, 
from $L_{\rm B}/M_b = 0.024 (L_{\rm B}/M)_\odot$ at $z=0$ 
to $\simeq 0.06 (L_{\rm B}/M)_\odot$ at $z=1$, though the error in each 
distant cluster is fairly large.
CL0939$+$47 deviates largely from this trend. 
We suspect, however, that the observed $L_{\rm B}/M_b$ of this cluster 
does not represent the real, global value, 
because the observed $L_{\rm B}/M_b$ is the value for
two substructures whose radii are only $r = 0.14 h^{-1}$ Mpc.
The CNOC clusters seem to have a larger scatter in $L_{\rm B}/M_b$.
This may reflect uncertainties 
due to a large factor of the conversion of 
optical luminosity from the value 
at the virial radius to that at $r=600 h^{-1}$ kpc.

% Fig. 6

There are two opposite explanations for the increase in $L_{\rm B}/M_b$ 
with redshift.
One is the brightening of $L_{\rm B}$ due to the luminosity 
evolution of the galaxy population as a whole.
Note that galaxy mergings, even if they occur, 
do not change the total mass of the galaxy population 
and that star formation which could be 
triggered by mergings can be treated 
in the framework of the {\lq}pure{\rq} luminosity evolution 
of the galaxy population.
The other explanation is that the mass of baryons ($\simeq$ hot gas) 
per galaxy decreases with redshift.
However, this explanation seems to be less plausible, 
because no significant evolution has been observationally 
found for the global properties of clusters at $z \ltsim 1$ 
(e.g., Schindler 1999)
(This result is, however, mainly for X-ray properties, and 
the evolution of the galaxy distribution in clusters is 
not well known).

In what follows, 
we take the former explanation as our hypothesis, i.e., 
we assume that the increase in $L_{\rm B}/M_b$ found here 
is due to pure brightening of the galaxy population as a whole.
Then the increase found here 
corresponds to brightening by $\simeq 1$ mag of the galaxy population.
In the next subsection, we compare the observed brightening 
with predictions of simple luminosity evolution models of galaxies.

%
%---- subsection 3.1 -----------------------------------------------
%

\subsection{Comparison with Luminosity Evolution Models}

We characterize the evolution of $L_{\rm B}$, 
the $B$-band luminosity summed over all galaxies in a cluster, 
by two parameters: 
the star formation timescale $\tau$ and the formation 
redshift $z_{\rm F}$.
In other words, we assume that all galaxies are formed at the same 
redshift $z_{\rm F}$ and that the $e$-folding time of star formation 
summed over all galaxies is $\tau$ (Gyr).
We compute $B$-band luminosity using the population synthesis 
code developed by Kodama and Arimoto (1997).
The values of $\tau$ and $z_{\rm F}$ examined here 
are $\tau = 0.1, 1, 2, 3$, and $5$ Gyr and $z_{\rm F} = 2, 3$, 
and $\infty$.
Models with $\tau = 0.1$ Gyr correspond to elliptical galaxies 
and models with $\tau = 5$ Gyr are for spiral disks like 
that of our Galaxy.
We do not examine $\tau > 5$ Gyr, since 
to take $\tau > 5$ Gyr leads to too blue colors at $z=0$ 
which are inconsistent with the observed colors 
of galaxies in nearby clusters  
(The mean $B-R_{\rm C}$ of the Virgo and Coma cluster galaxies is
$1.2$ and $1.8$, respectively 
[Andreon 1996 for Coma and Young and Currie 1998 for Virgo], 
while the models with $\tau=0.1$ and $\tau=5$ Gyr 
give $B-R_{\rm C}=1.6$ and $B-R_{\rm C}=1.1$, respectively, 
at an age of 12 Gyr).
We also set the lower limit of $z_{\rm F}$ to be 2 
following the traditional pure luminosity evolution models 
which assume that elliptical/S0 galaxies and spiral galaxies 
are formed at high redshifts ($z \gtsim 2$) 
and which broadly succeed in reproducing observed properties 
of these galaxies 
(e.g., Kodama et al. 1998; Shimasaku and Fukugita 1998).

Figure 6 compares the observed $L_{\rm B}/M_b$ with predictions.
Predicted values of $L_{\rm B}/M_b$ are normalized to match 
the observed value at $z=0$.
In other words, models are used to predict relative brightening 
(or fading) of $L_{\rm B}$ as a function of $z$.
Panels (a), (b), and (c) are for $z_{\rm F}=\infty, 3$, and 2, 
respectively.
From panel (a), we find that all the models reproduce the observation.
If, however, $z_{\rm F}=3$ is adopted (panel [b]), 
models with $\tau=1$ and 2 Gyr give too steep brightening 
compared with the observation.
This trend is strengthened for the $z_{\rm F}=2$ case (panel [c]):
the $\tau=0.1$ Gyr model also becomes inconsistent with the 
observation, 
though the discrepancy is at less than $2 \sigma$ levels.
The allowed range for $\tau$ is dependent on $z_{\rm F}$, 
and we cannot rule out any value of $\tau$ 
on the basis of the current data 
if we permit $z_{\rm F} = \infty$, 
though {\lq}single burst{\rq} models with $\tau =0.1$ Gyr and 
{\lq}disk{\rq} models with $\tau = 5$ (and $\tau=3$ Gyr models) 
match the observation for a wider range of $z_{\rm F}$ 
toward lower redshifts than the other ($\tau=1,2$ Gyr) models.

It is interesting that 
{\lq}single burst{\rq} ($\tau = 0.1$ Gyr) models and 
{\lq}disk{\rq} ($\tau = 5$ Gyr) models
are consistent with the observation.
In this paragraph, we concentrate on these models, 
and examine which are more 
consistent with the observed luminosity evolution of individual 
galaxies in clusters.
Various observations suggest that elliptical and S0 galaxies 
in clusters brighten by $\sim 1$ mag from $z=0$ to $z=1$, 
which is consistent with the single burst model 
(e.g., Schade, Barrientos, L\'{o}pez-Cruz 1997; 
Ellis et al. 1997; Kodama et al. 1998; van Dokkum et al. 1998).
Spiral galaxies have also been found to brighten by $\sim 1$ mag 
(e.g., Schade et al. 1996), though observations are not as many as 
those of elliptical and S0 galaxies.
The amounts of brightening of E/S0 and spiral galaxies are similar 
to each other, and they are in agreement with the predictions of 
$\tau=0.1$ Gyr models (with $z_{\rm F} \ge 3$) and $\tau=5$ Gyr models.
However, our distant clusters are rich clusters and thus 
it is likely that elliptical and S0 galaxies dominate in these 
clusters.
Hence, the $\tau=0.1$ Gyr models seem to be more plausible 
for describing the evolution of $L_{\rm B}/M_b$.
This is also supported by the fact that the mean color of 
galaxies in the Coma cluster, which is a very rich nearby 
cluster and is likely to be a counterpart of the distant clusters
studied here, agrees with the color predicted by 
the $\tau=0.1$ Gyr models.

Van Dokkum et al. (1998) present observations of 
$M/L$ ratio in the $B$ band of early-type galaxies 
in five clusters at $0.02 \le z \le 0.83$.
They find that the $M/L$ ratio evolves as 
$\Delta \log M/L_{\rm B} \propto -0.40 z$
($\Omega_0=0.3, \lambda_0=0$), 
which is consistent with single-burst models with 
$z_{\rm F}>1.7-2.8$.
If the evolution of $M/L$ found by van Dokkum et al. (1998) 
is understood as pure luminosity evolution of $L_{\rm B}$, 
the formula $\Delta \log M/L_{\rm B} \propto -0.40 z$  
implies that $L_{\rm B}$ brightens by a factor of 2.5 from $z=0$ to 1, 
which is in excellent agreement with the brightening of 
$L_{\rm B}/M_b$ found in this study.
Measurements of $M/L$ of individual galaxies are 
a direct measurement of the effects of luminosity evolution 
occurred in galaxies, while measurements of 
$L_{\rm B}/M_b$ of clusters are less direct.
Note, however, that information contained in $L_{\rm B}/M_b$ 
is different from that in $M/L$ of individual galaxies.
The quantity $L_{\rm B}/M_b$ describes the evolution of 
luminosity summed over {\it all} galaxies in clusters. 
$L_{\rm B}/M_b$ also gives us a hint to 
the star formation efficiency in clusters (see below).
In any case, the agreement of brightening between 
$M/L_{\rm B}$ and $L_{\rm B}/M_b$ found here 
can be regarded as indirect support of our conclusion that 
single-burst like models seem to be plausible 
for describing the evolution of $L_{\rm B}/M_b$.

The absolute value of $L_{\rm B}/M_b$ tells us a hint to 
the star formation efficiency in clusters, 
i.e., the fraction of baryons in clusters used to form stars.
If the star formation efficiency differs among clusters, 
the absolute value of $L_{\rm B}/M_b$ would also vary 
from cluster to cluster.
The fact that there exist models, such as those with $\tau=0.1$ Gyr, 
which reproduce the observed $L_{\rm B}/M_b$ of many clusters 
at different redshifts within the observational errors 
suggest that the star formation efficiency is universal 
among clusters up to $z \sim 1$.

% Fig. 7

In order to see when baryons are converted into stars, 
we plot in figure 7 the predicted evolution of $M_\star/M_b$, 
where the evolution of 
$M_\star$ is calculated from mass-to-luminosity ratios of 
galaxies predicted by the $z_{\rm F}=3$ models. 
The values of $M_\star/M_b$ at $z=0$ have been normalized so that they 
are consistent with the observed $L_{\rm B}/M_b$ 
of the nearby clusters
\footnote{$M_\star/M_b(z=0) = (M/L_{\rm B})_\star^{\rm pred}(z=0) 
\times (L_{\rm B}/M_b)^{\rm obs}(z=0)$}.
The predicted values of $M_\star/M_b$ at $z=0$ are $0.05-0.13$, 
depending on $\tau$, and are consistent with 
the observed value (the filled square with an error bar). 
This simply implies that the predicted $(M/L_{\rm B})_\star$ values at 
$z=0$ fall within $(6 \pm 3)h (M/L_{\rm B})_\odot$, 
which is the adopted value of
$(M/L_{\rm B})_\star$ for galaxies in nearby clusters.
As expected, the evolution of $M_\star/M_b$ largely differs 
among the models.
A constant $M_\star/M_b$ is predicted by the $\tau=0.1$ Gyr model 
in the redshift range of this figure,
while for the $\tau=5$ Gyr model, 
half of the stars present today are formed at $z < 1$.

Finally, we mention how to put stronger constraints on models 
using the evolution of $L_{\rm B}/M_b$.
Unfortunately, the steepness of brightening up to $z=1$ 
is not a monotonic function of $\tau$: the brightening is the steepest 
for $\tau=1-2$ Gyr models, and models with 
$\tau=0.1$ Gyr and $\tau=3-5$ Gyr give similar brightening.
In order to place further constraints 
using the evolution of $L_{\rm B}/M_b$,
one needs data at $z>1$: for example, 
for $z_{\rm F}=3$ and 2, $\tau=0.1$ Gyr models predict much steeper 
brightening at $z > 1$ than $\tau = 5$ Gyr models.

%
%---- subsection 3.2 -----------------------------------------------
%

\subsection{Comparison with Evolution of Field Galaxies}

The quantity $L_{\rm B}/M_b$ is the $B$-band luminosity per unit baryon 
in clusters.
The corresponding quantity for field galaxies is 
the ratio of the blue luminosity density 
$l_{\rm B}$ [${L_{\rm B}}_\odot$ Mpc$^{-3}$] to 
the mean baryon density $\rho_{\rm b}$ [$\MO$ Mpc$^{-3}$].
In this subsection, we derive $l_{\rm B}/\rho_b$ of field galaxies 
up to $z \sim 1$ from 
the literature and compare it with $L_{\rm B}/M_b$ of clusters.

Data of $l_{\rm B}$ are taken (or computed) from recent measurements 
of luminosity function based on redshift surveys: 
Lilly et al. (1996; CFRS; data points are at $z=0.35, 0.625, 0.875$), 
Ellis et al. (1996; Autofib; $z=0.085, 0.25, 0.55$), 
Colless (1998; 2dF; $z=0.11$), 
Loveday et al. (1992; APM; $z \simeq 0.05$), 
Zucca et al. (1997; ESP; $z \simeq 0.1$), and
Marzke et al. (1998; SSRS2; $z \simeq 0.025$).
Lilly et al. (1996) give $l_{\rm B}$ itself while the other papers give 
only luminosity functions in the $B$ band.
For those except for Lilly et al. (1996), 
we integrate the luminosity function given in each paper 
from $M_{\rm B}=-25$ to $-10$ to obtain $l_{\rm B}$.

In order to compute $\rho_{\rm b}$,
we adopt $\Omega_b = 0.02 h^{-2}$ following 
Tytler and his coworkers' results 
(e.g., Burles, Tytler 1998). 
Their estimates of $\Omega_b$ 
are based on measurements of deuterium abundance (D/H) of 
QSO absorption systems at high redshifts. 
Note that the measurements of $\Omega_b$ have not completely 
converged among authors, ranging from $\Omega_b h^2 \simeq 0.01$ 
to 0.02, though Tytler et al.'s 
measurements seem to be the most reliable (e.g., Turner 1999).

% Fig. 8

The $l_{\rm B}/\rho_b$ of field galaxies calculated above are 
plotted in figure 8 as open circles with error bars.
A gradual increase in $l_{\rm B}/\rho_b$ with redshift is seen.
This is due to the brightening of $l_B$
\footnote{
For field galaxies, the global star formation 
rate ($\dot{\rho}_\star$ [$\MO$ yr$^{-1}$ Mpc$^{-3}$]) 
has also been measured from UV and emission-line luminosities 
of galaxies (e.g., Madau et al. 1998; for recent observations, 
see Cowie, Songaila, Barger 1999).
These measurements suggest that $\dot{\rho}_\star$ increases 
by a factor of $\sim 3-10$ 
from $z=0$ to $1$, but the scatter among the data is still large.}.

A fact which needs attention is 
that the values of $l_{\rm B}/\rho_b$ at $z \ltsim 0.05$ 
are smaller than those at $z \simeq 0.1$ 
by as large as a factor of $\simeq 2$.
It is unlikely that the luminosity density evolves so rapidly 
for such a short time from $z=0.1$ to the present.
A possible explanation for this problem  
is that the number density of galaxies in the local 
($z \ltsim 0.1$) universe happens to be lower than the global 
value (e.g., Marzke et al. 1998), 
though further investigations are needed 
in order to prove or disprove this explanation.
In this paper, we regard 
the values at $z \simeq 0.1$ as the local ($z=0$) value.

%
%---- subsubsection 3.2.1 ------------------------------------------
%

\subsubsection{Local values for $L_{\rm B}/M_b$ and $l_{\rm B}/\rho_b$}

The filled circles in figure 8 indicate the $L_{\rm B}/M_b$ of 
the distant clusters of galaxies.
The filled square at $z=0$ corresponds 
to the average of the nearby clusters.
We find that the local $L_{\rm B}/M_b$ agrees 
with the local $l_{\rm B}/\rho_b$: 
$L_{\rm B}/M_b$ is $(0.024 \pm 0.004) (L_{\rm B}/M)_\odot$ 
and the average of the three $l_{\rm B}/\rho_b$ values 
at $z \simeq 0.1$ is $(0.026 \pm 0.002) (L_{\rm B}/M)_\odot$
\footnote{
A note should be added.
The agreement found here does not hold
if very different values for $\Omega_b$ and $h$ are adopted, 
because 
(i) $l_{\rm B}/\rho_b$ changes linearly with $\Omega_b^{-1}$ 
and 
(ii) the dependence of $l_{\rm B}/\rho_b$ and $L_{\rm B}/M_b$ 
on $h$ is different ($l_{\rm B}/\rho_b \propto h$ 
if $\Omega_b h^2$ is fixed, 
while $L_{\rm B}/M_b \propto h^{0.5}$), 
though we think that the values of $\Omega_b$ and $h$ 
adopted in this paper are the most probable at present.}.

This agreement implies that the blue luminosity per unit baryon mass 
is very close between clusters and fields. 
Then the next question may be whether the stellar mass per 
unit baryon mass ($M_\star/M_b$), 
i.e., the star formation efficiency, is
the same between clusters and fields 
(Note that morphological type mix largely differs between 
clusters and fields and that $(M/L_{\rm B})_\star$ of galaxies
varies with morphology).
In order to examine this, we do a simple 
(but more detailed than that given in \S 2.1) 
estimation of $M_\star/M_b$ of clusters and fields below.
We assume the $(M/L_{\rm B})_\star$ of elliptical and S0 
galaxies to be $8h (M/L_{\rm B})_\odot$ and that of spiral 
and irregular galaxies to be $4h (M/L_{\rm B})_\odot$
\footnote{The value for elliptical and S0 galaxies is based on 
van der Marel's (1991) study, which found 
that the $M/L_{\rm B}$ of elliptical galaxies with 
$1 \times 10^{10} h^{-2} {L_{\rm B}}_\odot$ 
is $8.4h (M/L_{\rm B})_\odot$.
The value for the Galactic disk, 
$M/L_{\rm B} \simeq 3 (M/L_{\rm B})_\odot$, is adopted 
as the value for spiral and irregular galaxies 
($4h \simeq 3$ for $h=0.7$).}.
Using these values and taking account of the mean 
morphological type mix of the 12 clusters, 
we obtain $M_\star/M_b = 0.114 \pm 0.020$.
A similar calculation for the field galaxies 
gives $M_\star/M_b = 0.097 \pm 0.009$ 
(We use the type mix given in Colless 1998:
$l_{\rm B}({\rm E+S0})/l_{\rm B}({\rm tot})=34.5\%$ and 
$l_{\rm B}({\rm S + Irr})/l_{\rm B}({\rm tot})=65.5\%$).
Hence, the $M_\star/M_b$ of clusters is in agreement with that in 
fields within errors.
Renzini (1997) has also obtained a similar result 
from a rough calculation for $M_\star/M_b$ of the Coma cluster 
and field galaxies.

%
%---- subsubsection 3.2.2 ------------------------------------------
%

\subsubsection{Evolution of $L_{\rm B}/M_b$ and $l_{\rm B}/\rho_b$}

From figure 8, it is found that the evolution of 
$L_{\rm B}/M_b$ as a function of redshift 
is the same as for $l_{\rm B}/\rho_b$
within the observational errors.
Both {\lq}brighten{\rq} by a factor of $2-3$ from $z=0$ to $z=1$.

The agreement between the evolution of 
$L_{\rm B}/M_b$ and $l_{\rm B}/\rho_b$ may be interpreted as 
the mean star formation history of cluster galaxies 
being similar to that of field galaxies.
However, we think that this agreement is probably superficial.
The observed global star formation rate of field galaxies 
increases by a factor of $\sim 3-10$ from the present epoch to 
$z = 1$, and then has a peak at $z=1-2$ 
(e.g., Madau et al. 1998; Cowie et al. 1999).
Though this star formation history, which reproduces 
the observed evolution of $l_{\rm B}/\rho_b$ as well, should 
be a solution for the mean star formation history of cluster galaxies,  
quite different models can also be solutions, 
as has been seen in \S 3.1.
Simple models having just two parameters 
($\tau$ and $z_{\rm F}$) were examined in \S 3.1, and many models 
have been found to reproduce the observed evolution of $L_{\rm B}/M_b$, 
and the color of galaxies in nearby rich clusters 
suggests that $\tau = 0.1$ Gyr models are favored (see \S 3.1).

In any case, 
we cannot give a clear conclusion about the mean star formation 
history of cluster galaxies on the basis of the current data.
As was mentioned in \S 3.1, data at $z > 1$ are useful 
to place further constraints on the star formation history.
More desirable may be data of ultraviolet luminosity 
per unit baryon mass, $L_{\rm UV}/M_b$,
from which one can measure star formation rate 
(and efficiency) directly.

%
%---- section 4 ----------------------------------------------------
%

\section{Conclusions}

We have derived $L_{\rm B}/M_b$ for massive 
($M_{\rm gas}$ at the Abell radius is 
$\ge 1 \times 10^{13} h^{-2.5} \MO$) clusters of galaxies 
up to $z \simeq 1$ from optical and X-ray data in the literature. 
Twenty-two clusters in our sample are at $z > 0.1$.
Assuming that the relative mix of hot gas and galaxies in clusters
does not change (i.e., no segregation in hot gas or galaxies)  
during cluster evolution, 
we use $L_{\rm B}/M_b$ to probe the star formation history 
of the galaxy population as a whole in clusters. 
We have found that the $L_{\rm B}/M_b$ of clusters increases 
with redshift from $L_{\rm B}/M_b=0.024 (L_{\rm B}/M)_\odot$ ($z=0$) 
to $\simeq 0.06 (L_{\rm B}/M)_\odot$ ($z=1$), 
indicating a factor of $\sim 2-3$ brightening.
This amount of brightening is almost identical to 
the brightening of the $M/L_{\rm B}$ ratio of early-type galaxies 
in clusters at $0.02 \le z \le 0.83$ reported by 
van Dokkum et al. (1998).

We have compared this result with luminosity evolution models 
for the galaxy population as a whole 
by changing the $e$-folding time of star formation $\tau$ by 
$0.1 \le \tau \le 5$ Gyr  
and the formation redshift $z_{\rm F}$ by $2 \le z_{\rm F} < \infty$.
We have found that {\lq}single burst{\rq} 
models ($\tau =0.1$ Gyr models) with $z_{\rm F} \ge 3$ 
and {\lq}disk{\rq} models ($\tau = 5$ Gyr) 
with arbitrary $z_{\rm F}$ are 
consistent with the observed brightening of blue luminosity 
to $z=1$, while models with $1 \le \tau \le 2$ Gyr 
tend to predict too steep brightening 
though we cannot rule out these models.

We have also derived the ratio of blue luminosity density to 
baryon density, $l_{\rm B}/\rho_b$, for field galaxies 
up to $z \simeq 1$ from various existing data, 
adopting $\Omega_b h^2 = 0.02$, 
and have found that 
the observed evolution of $L_{\rm B}/M_b$ agrees with  
that of $l_{\rm B}/\rho_b$, including the absolute values, 
from the present epoch to $z \simeq 1$ 
within the observational uncertainties, 
indicating that blue luminosity per unit baryon mass is 
similar between clusters and fields up to $z \simeq 1$.
We have made a simple estimate of star formation efficiency 
($M_\star/M_b$) to find no difference between clusters and fields.
To place further constraints on the mean star formation 
history of cluster galaxies needs new data at higher redshifts 
or direct measurements of star formation rate.

%
%------------------------------------------------------------------
%

\vspace{10pt}
\noindent
Acknowledgements

\noindent
We thank the anonymous referee for useful comments which 
improved the paper.
This research was supported in part by 
the Grants-in-Aid by the Ministry of Education, 
Science, Sports and Culture of Japan (07CE2002) 
to RESCEU (Research Center for the Early Universe).

%
%------------------------------------------------------------------
%

\section*{References}
 
\re
Andreon, S. 1996, A\&A 314, 763
 
\re
Arnaud, M., Rothenflug, R., Boulade, O., Vigroux, L., 
Vangioni-Flam, E. 
1992, A\&A 254, 49
 
\re
Bahcall, J. N.
1984, ApJ 287, 926
 
\re
Balogh, M., Schade, D., Morris, S. L., Yee, H. K. C., 
Carlberg, R. G., Ellingson, E.
1998, ApJ 504, L75
 
\re
Benitez, N., Broadhurst, T., Rosati, P., Courbin, F., Squires, G., 
Lidman, C., Magain, P. 
1998, astro-ph/9812218
 
\re
Broeils, A. H., Couteau, S.
1996,
in {\it Dark and Visible Matter in Galaxies}, 
ASP Conference Series, Vol. 117; 1997; 
ed. M. Persic and P. Salucci, p74
 
\re
Burles, S., Tytler, D.
1998, ApJ 507, 732
 
\re
Carlberg, R. G., Yee, H. K. C., Ellingson, E., Abraham, R., 
Gravel, P., Morris, S., Pritchet, C. J.
1996, ApJ 462, 32
 
\re
Cirimele, G., Nesci, R., Tr\`{e}vese, D. 1997, ApJ 475, 11
 
\re
Colless, M. 1998, astro-ph/9804079
 
\re
Cowie, L. L., Songaila, A., Barger, A. J. 
1999, astro-ph/9904345
 
\re
Dressler, A., Oemler, A., Jr., Couch, W. J., Smail, I., 
Ellis, R. S., Barger, A., Butcher, H., Poggianti, B. M., 
et al. 
1997, ApJ 490, 577
 
\re
Ellis, R. S., Colless, M., Broadhurst, T., Hely, J., 
Glazebrook, K. 1996, MNRAS 280, 235
 
\re
Ellis, R. S., Smail, I., Dressler, A., Couch, W. J., Oemler, A., Jr., 
Butcher, H., Sharples, R. M. 1997, ApJ 483, 582
 
\re
Fischer, P., Tyson, A. J.
1997, AJ 114, 14
 
\re
Freedman, W. L. 1999, astro-ph/9905222
 
\re
Fukugita, M., Shimasaku, K., Ichikawa, T.
1995, PASP 107, 945
 
\re
Hattori, M., Ikebe, Y., Asaoka, I., Takeshima, T., 
B\"ohringer, H., Mihara, T., Neumann, D. M., 
Schindler, S., Tsuru, T., Tamura, T.
1997, Nature 388, 146
 
\re
Infante, L., Fouque, P., Hertling, G., Way, M. J., 
Giraud, E., Quintana, H.
1994, A\&A 289, 381
 
\re
Kodama, T., Arimoto, N. 1997, A\&A 320, 41
 
\re
Kodama, T., Arimoto, N., Barger, A. J., Arag\'on-Salamanca, A.
1998, A\&A 334, 99
 
\re
Lewis, A. D., Ellingson, E., Morris, S. L., Carlberg, R. G.
1999, ApJ 517, 587
 
\re
Lilly, S. J., Le F\`evre, O., Hammer, F., Crampton, D.
1996, ApJ 460, L1
 
\re
Lilly, S. J., Schade, D., Ellis, R., Le F\`evre, O., 
Brinchmann, J., Tresse, L., Abraham, R., Hammer, F.,
et al. 
1998, ApJ 500, 75
 
\re
Loveday, J., Peterson, B. A., Efstathiou, G., Maddox, S. J.
1992, ApJ 390, 338
 
\re
Madau, P., Pozzetti, L., Dickinson, M.
1998, ApJ 498, 106
 
\re
Marzke, R. O., da Costa, L. N., Pellegrini, P. S., 
Willmer, C. N. A., Geller, M. J. 
1998, ApJ 503, 617
 
\re
Neumann, D. M., B\"ohringer, H.
1997, MNRAS 289, 123
 
\re
Renzini A. 
1997, ApJ 488, 35
 
\re
Pizzella, A., Amico, P., Bertola, F., Buson, L. M., 
Danziger, I. J., Dejonghe, H., Sadler, E. M., Saglia, R. P., 
de Zeeuw, P. T., \& Zeilinger, W. W.
1997, A\&A 323, 349
 
\re
Poggianti, B. M., Smail, I., Dressler, A., Couch, W. J., 
Barger, A., Butcher, H., Ellis, R. S., Oemler, A., Jr., 
1999, ApJ 518, 576
 
\re
Sahu, K. C., Shaw, R. A., Kaiser, M. E., Baum, S. A., 
Ferguson, H. C., Hayes, J. J. E., Gull, T. R., Hill, R. J., 
Hutchings, J. B., Kimble, R. A., Plait, P., \& Woodgate, B. E.
1998, ApJ 492, L125
 
\re
Schade, D., Carlberg, R. G., Yee, H. K. C., L\'{o}pez-Cruz, O., 
Ellingson, E.
1996, ApJ 465, L103
 
\re
Schade, D., Barrientos, L. F., L\'{o}pez-Cruz, O.
1997, ApJ 477, L17
 
\re
Schindler, S.
1999, astro-ph/9908130
 
\re
Schindler, S., Belloni, P., Ikebe, Y., Hattori, M., 
Wambsganss, J., Tanaka, Y.
1998, A\&A 338, 843
 
\re
Schindler, S., Wambsganss, J. 
1997, A\&A 322, 66
 
\re
Shimasaku, K., Fukugita, M.
1998, ApJ 501, 578
 
\re
Squires, G., Kaiser, N., Babul, A., Fahlman, G., 
Woods, D., Neumann, D. M., B\"ohringer, H. 
1996, ApJ 461, 572
 
\re
Squires, G., Neumann, D. M., Kaiser, N., Arnaud, M., 
Babul, A., B\"ohringer, H., Fahlman, G., Woods, D.
1997, ApJ 482, 648
 
\re
Taguchi, H., Shimasaku, K., Doi, M., Okamura, S.
1999, in preparation
 
\re
Turner, M. S. 1999, astro-ph/9904051
 
\re
van der Marel, R. P. 
1991, MNRAS 253, 710
 
\re
van Dokkum, P. G., Franx, M., Kelson, D. D., Illingworth, G. D.
1998, ApJ 504, L17
 
\re
Vogt, N. P., Phillips, A. C., Faber, S. M., Gallego, J., 
Gronwall, C., Guzm\'an, R., Illingworth, D., Koo, D. C., 
\& Lowenthal, J. D. 
1997, ApJ 479, L121
 
\re
Young, C. K.,  Currie, M. J. 1998, A\&AS 127, 367
 
\re
Zucca, E., Zamorani, G., Vettolani, G., Cappi, A., 
Merighi, R., Mignoli, M., Stirpe, G. M., MacGillivray, H., 
et al. 
1997, A\&A 326, 477
 
\newpage

%-------------------------------------------------------------------

Table~1.\hspace{4pt}Distant Clusters. 

\vspace{20pt}
\begin{tabular}{lccrrcc}
\hline
\hline
name  &   $z$  & $L_{\rm B}/M_b^{a)}$ & $L_{\rm B}^{b)}$ 
& $M_{\rm gas}^{c)}$ 
& radius$^{d)}$  & error$^{e)}$ \\
\hline
A1413         & $0.14$ & $0.028$ &  7.6  & 2.42  &  0.77   & $21$  \\
A1689         & $0.18$ & $0.042$ & 14.0  & 2.80  &  0.78   & $21$  \\
A2218         & $0.18$ & $0.043$ &  8.4  & 1.81  &  0.42   & $33$  \\
A2163         & $0.20$ & $0.037$ &  3.0  & 0.74  &  0.26   & $48$  \\
CL0500-24     & $0.32$ & $0.037$ &  3.9  & 0.95  &  0.53   & $30$  \\
CL0939+47     & $0.41$ & $0.089$ &  1.8  & 0.18  &  0.14   & ---   \\
RXJ1347.5-1145& $0.45$ & $0.053$ & 65.7  & 11.3  &  1.09   & $37$  \\
CL0016+16     & $0.55$ & $0.044$ & 50.6  & 10.4  &  1.67   & ---   \\
AXJ2019+1127  & $1.01$ & $0.075$ &  5.6  & 0.68  &  0.30   & $30$  \\
\hline
    A2390     & 0.23   & 0.030   & 11.3  & 3.3   & 0.60 & --- \\
MS0302+16     & 0.42   & 0.055   &  5.1  & 0.84  & 0.60 & --- \\
MS0440+02     & 0.20   & 0.027   &  2.5  & 0.84  & 0.60 & --- \\
MS0451+02     & 0.20   & 0.021   &  6.3  & 2.7   & 0.60 & --- \\
MS0451-03     & 0.54   & 0.036   & 14.7  & 3.7   & 0.60 & --- \\
MS0839+29     & 0.19   & 0.054   &  4.5  & 0.74  & 0.60 & --- \\
MS0906+11     & 0.17   & 0.066   &  9.2  & 1.3   & 0.60 & --- \\
MS1006+12     & 0.26   & 0.028   &  6.4  & 2.0   & 0.60 & --- \\
MS1008-12     & 0.31   & 0.051   & 11.2  & 2.0   & 0.60 & --- \\
MS1224+20     & 0.33   & 0.066   &  7.1  & 0.98  & 0.60 & --- \\
MS1358+62     & 0.33   & 0.062   & 11.4  & 1.7   & 0.60 & --- \\
MS1455+22     & 0.26   & 0.021   &  4.9  & 2.1   & 0.60 & --- \\
MS1512+36     & 0.37   & 0.028   &  4.2  & 1.4   & 0.60 & --- \\
\hline
\end{tabular}

\vspace{6pt}\par\noindent
a) In units of $h^{0.5} (L_{\rm B}/M)_\odot$.
\par\noindent
b) In units of $h^{-2} \times 10^{11} \LO$.
\par\noindent
c) In units of $h^{-2.5} \times 10^{13} \MO$.
\par\noindent
d) Radius in units of $h^{-1}$ Mpc adopted to measure 
$L_{\rm B}$ and $M_{\rm gas}$.
\par\noindent
e) Relative error ($\%$).

%
%------------------------------------------------------------------
%

\clearpage
\section*{Figure Captions}

\noindent
Fig.1. $L_{\rm V}$ plotted against $M_{\rm gas}$ for 
Arnaud et al.'s (1992) nearby clusters.
The filled and open circles indicate 
clusters with and without type mix data, respectively.
The dashed and solid lines indicate the best fit of a power law 
to all and massive ($M_{\rm gas} \ge 1 \times 10^{13} h^{-2.5} \MO$) 
clusters, respectively. 
The dotted line corresponds to the best fit of 
$L_{\rm V} \propto M_{\rm gas}$ to the massive clusters.

\vspace{10pt}
\noindent
Fig.2. $L_{\rm V}/M_{\rm gas}$ plotted against 
the fraction of luminosity emitted from elliptical and S0 galaxies
to the total luminosity $L_{\rm V}({\rm E/S0})/L_{\rm V}$ 
for Arnaud et al.'s (1992) clusters having type mix data.
The filled and open circles indicate 
$M_{\rm gas} \ge 1 \times 10^{13} h^{-2.5} \MO$ and 
$M_{\rm gas} < 1 \times 10^{13} h^{-2.5} \MO$ clusters, respectively.

\vspace{10pt}
\noindent
Fig.3. $L_{\rm B}/M_b$ as a function of $M_{\rm gas}$ for 
Arnaud et al.'s (1992) clusters.
The filled and open circles indicate 
clusters with and without type mix data, respectively.

\vspace{10pt}
\noindent
Fig.4. $L_{\rm B}$ versus $M_{\rm gas}$ for the 12 nearby and 
the 22 distant clusters adopted in this paper.
Filled circles indicate the nearby clusters.
Open circles and crosses are for $0.1 < z \le 0.4$ and 
$0.4 < z \le 0.7$ clusters, respectively.
The star corresponds to the most distant cluster 
AXJ2019+1127 at $z=1.01$. 
The thick solid line indicates the best fit of 
$L_{\rm B} \propto M_{\rm gas}$ to the nearby clusters.
The thin solid line and the dotted line correspond to 
a similar fit to the distant clusters at $0.1 < z \le 0.4$ and 
$0.4 < z \le 0.7$, respectively.

\vspace{10pt}
\noindent
Fig.5. Dependence of the measurement of $L_{\rm B}/M_{\rm gas}$ 
on $\Omega_0$ and $\lambda_0$.
The ratio of $L_{\rm B}/M_{\rm gas}(\Omega_0,\lambda_0)$
to $L_{\rm B}/M_{\rm gas}(0.2,0)$ is 
plotted as a function of redshift 
for $(\Omega_0,\lambda_0)=(1,0)$ and $(0.2, 0.8)$ cases.

\vspace{10pt}
\noindent
Fig.6. Observed $L_{\rm B}/M_b$ of clusters as a function of redshift.
The filled circles present the 22 distant clusters and 
the filled square indicates the average $L_{\rm B}/M_b$ of the nearby 
clusters.
Model predictions are overlaid.
Predicted values of $L_{\rm B}/M_b$ are normalized to match 
the observed value at $z=0$.
Panels (a), (b), and (c) are 
for $z_{\rm F}=\infty, 3$, and 2, respectively.
Thick and thin solid lines indicate models with $\tau=0.1$ and 
1 Gyr, respectively.
Dotted, dashed, and long-dashed lines correspond to 
models with $\tau = 2, 3$, and 5 Gyr, respectively.

\vspace{10pt}
\noindent
Fig.7. The ratio of stellar mass to baryon mass, plotted as a function 
of redshift. 
The five lines indicate predictions of models with different $\tau$.
All models are for $z_{\rm F}=3$.
The meanings of lines are the same as in figure 6.
The filled square corresponds to the observed value.

\vspace{10pt}
\noindent
Fig.8. $l_{\rm B}/\rho_b$ of field galaxies as a function of redshift, 
compared with $L_{\rm B}/M_b$ of clusters.
The filled circles indicate the 22 distant clusters and 
the filled square is for the average $L_{\rm B}/M_b$ of the nearby 
clusters.
The open circles with error bars correspond to $l_{\rm B}/\rho_b$ 
of field galaxies.

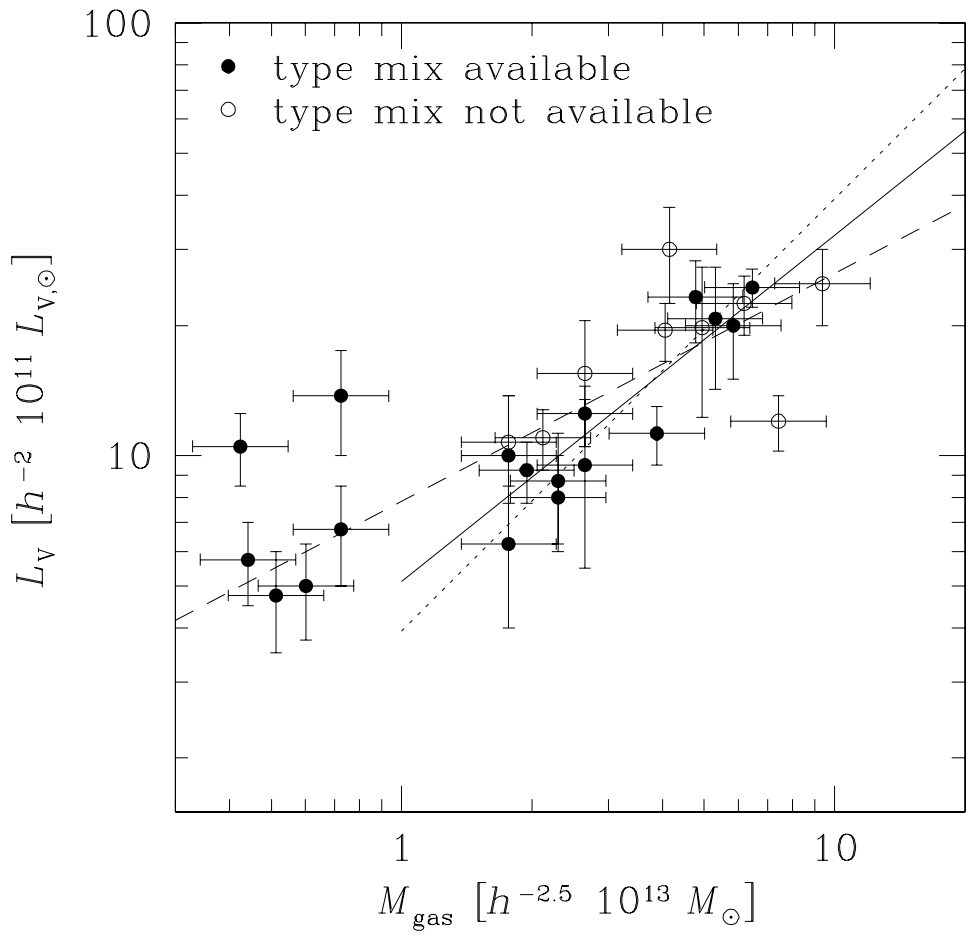
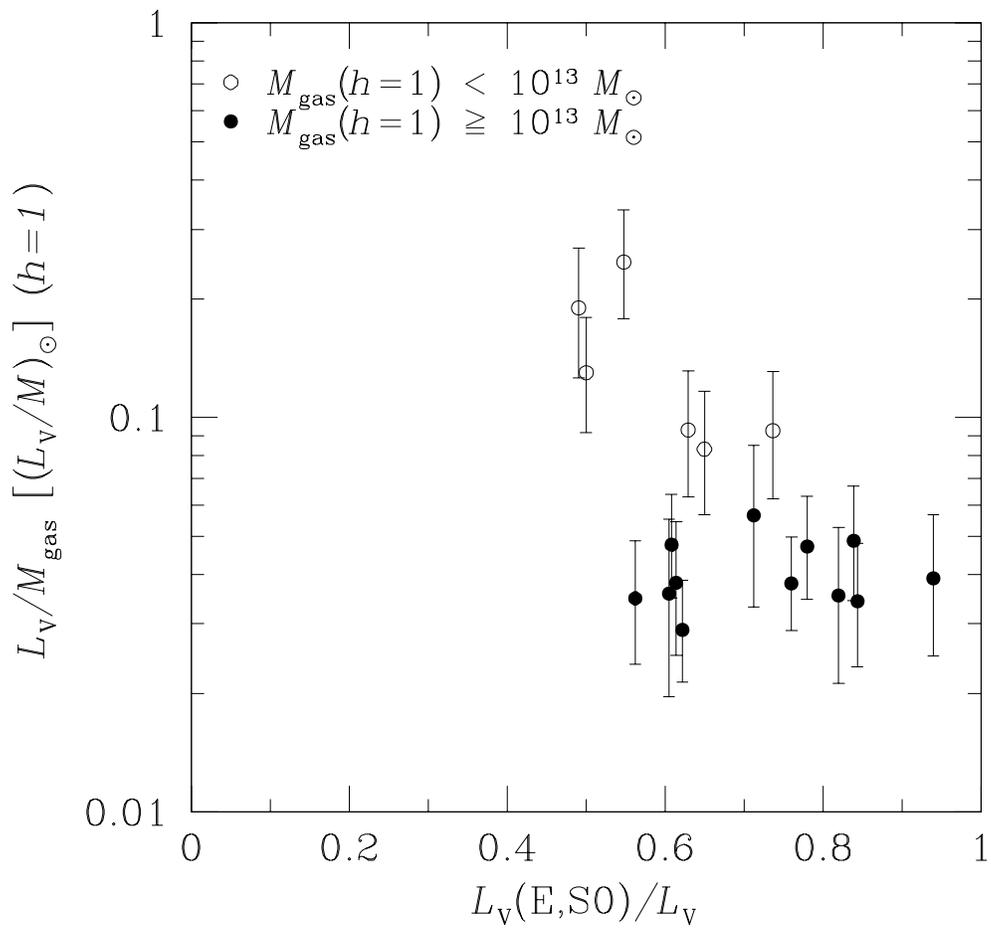
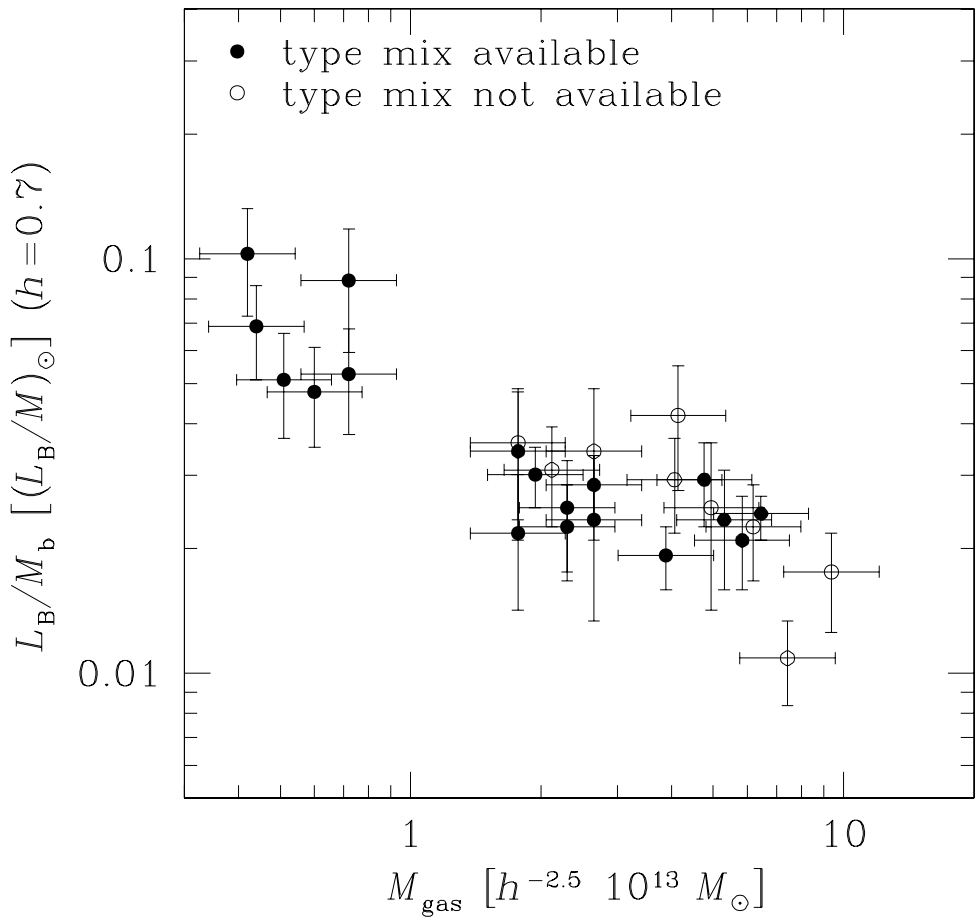
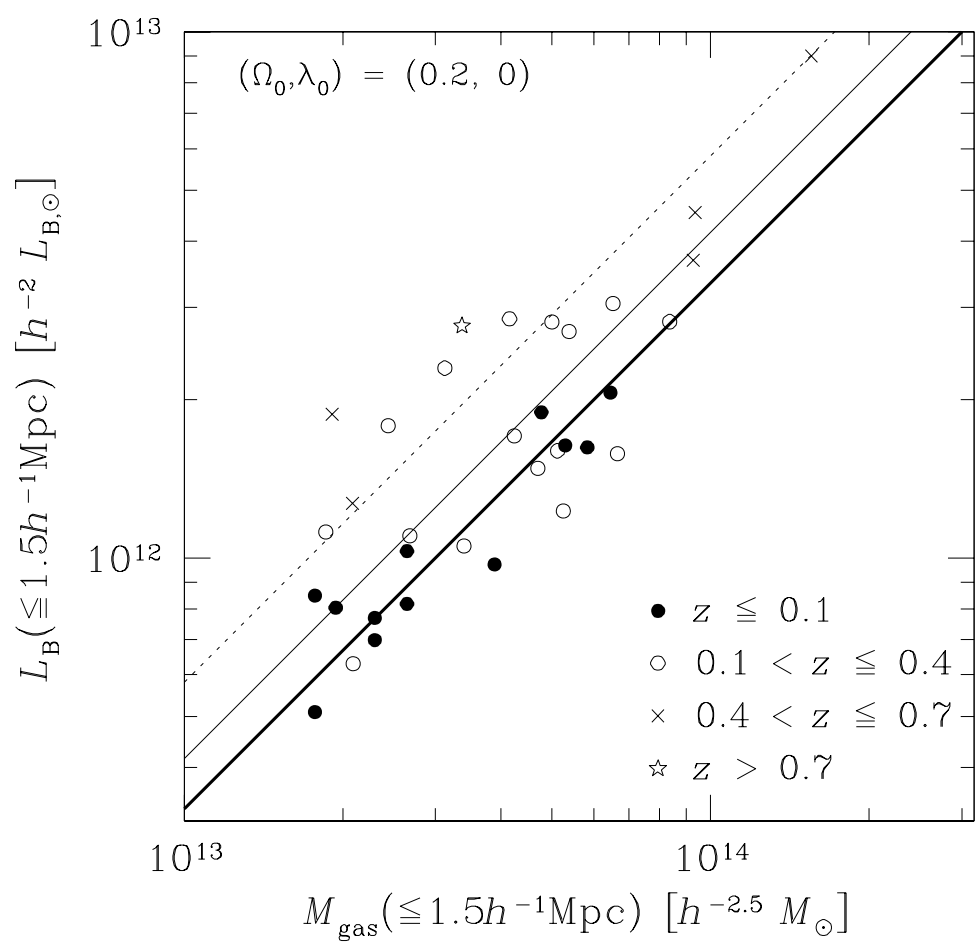
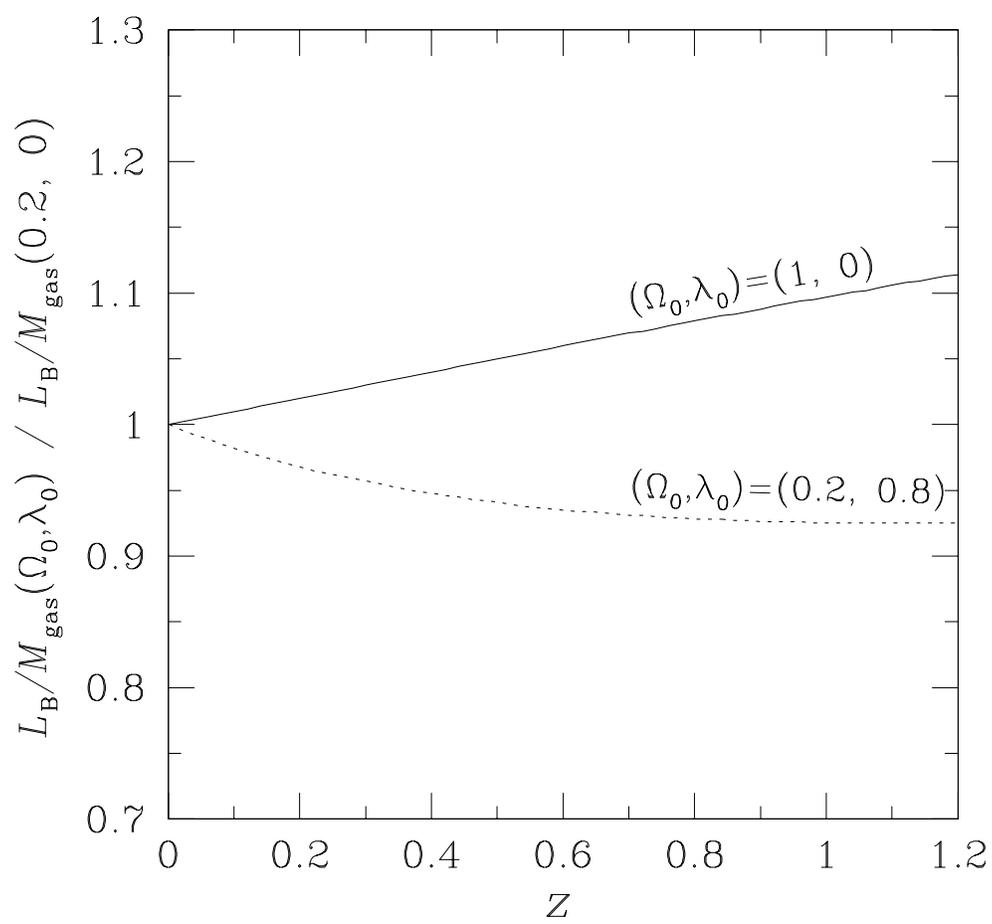
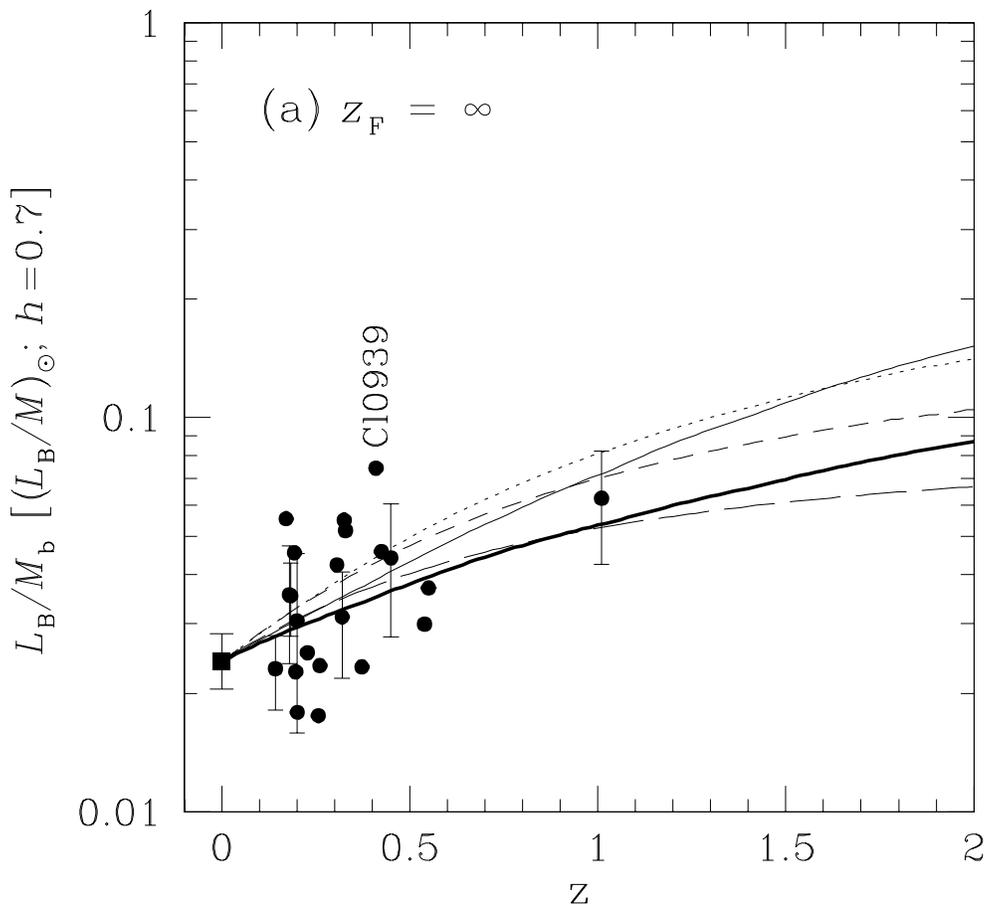
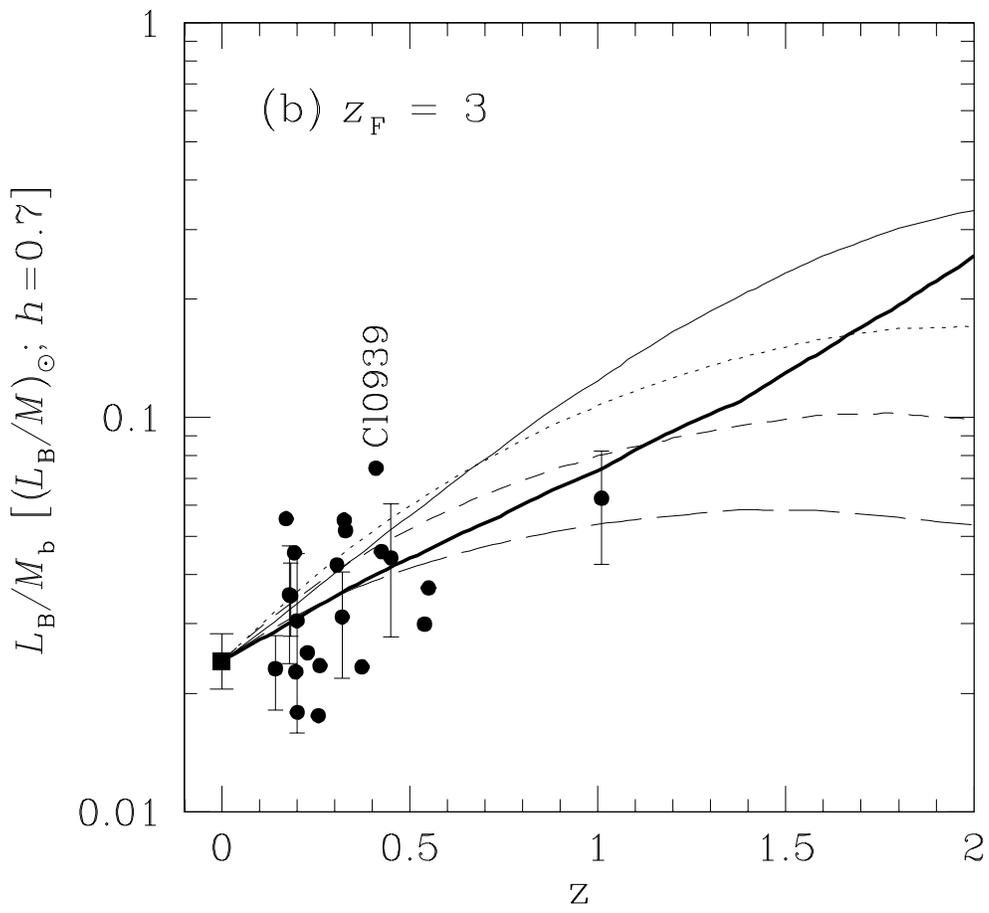
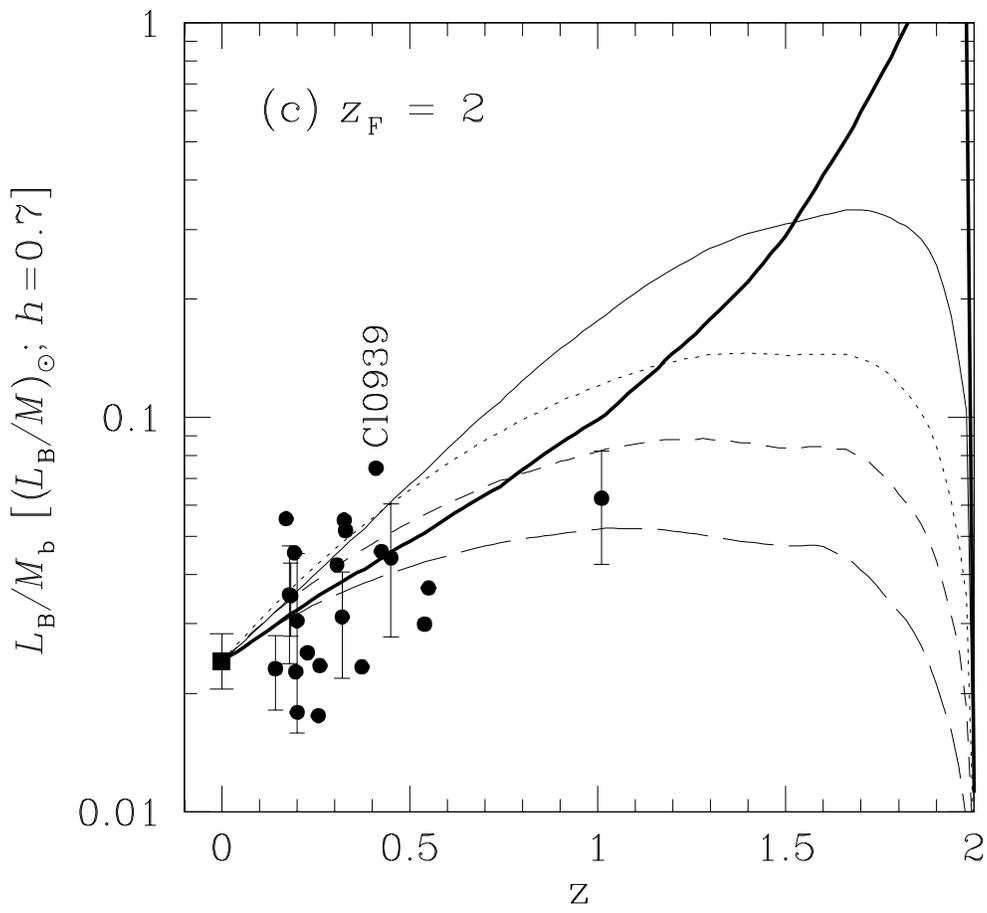
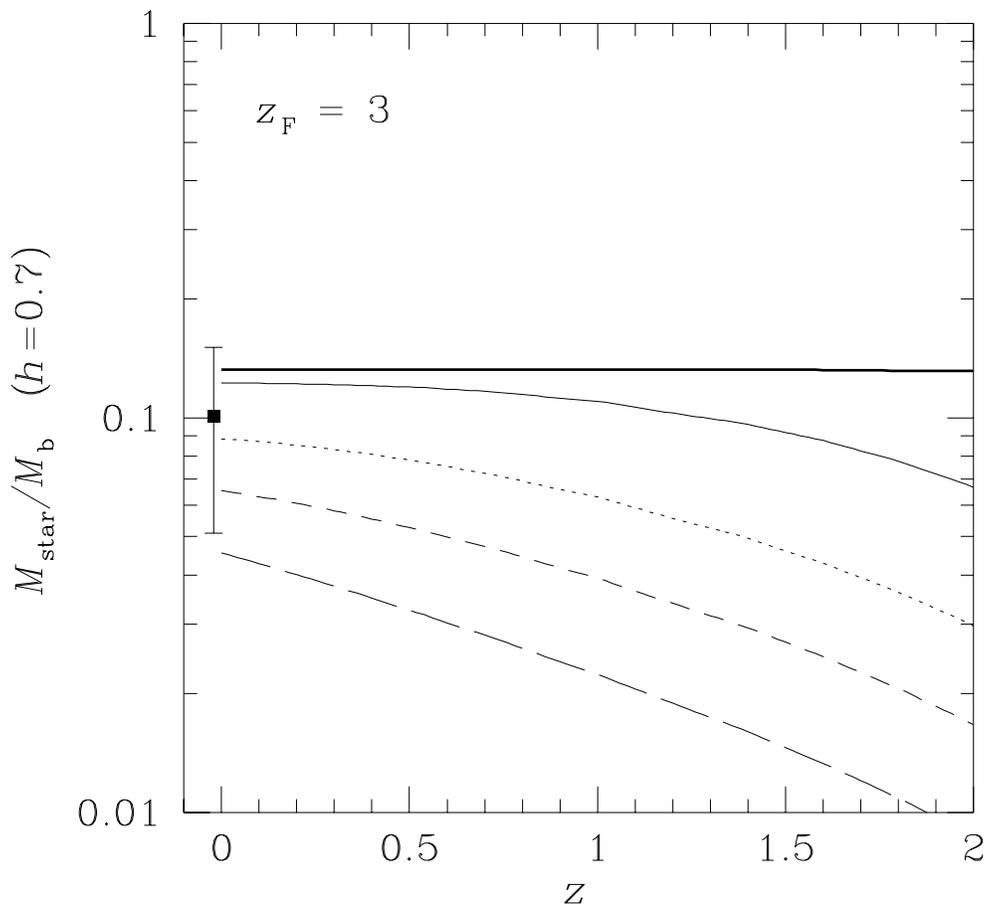
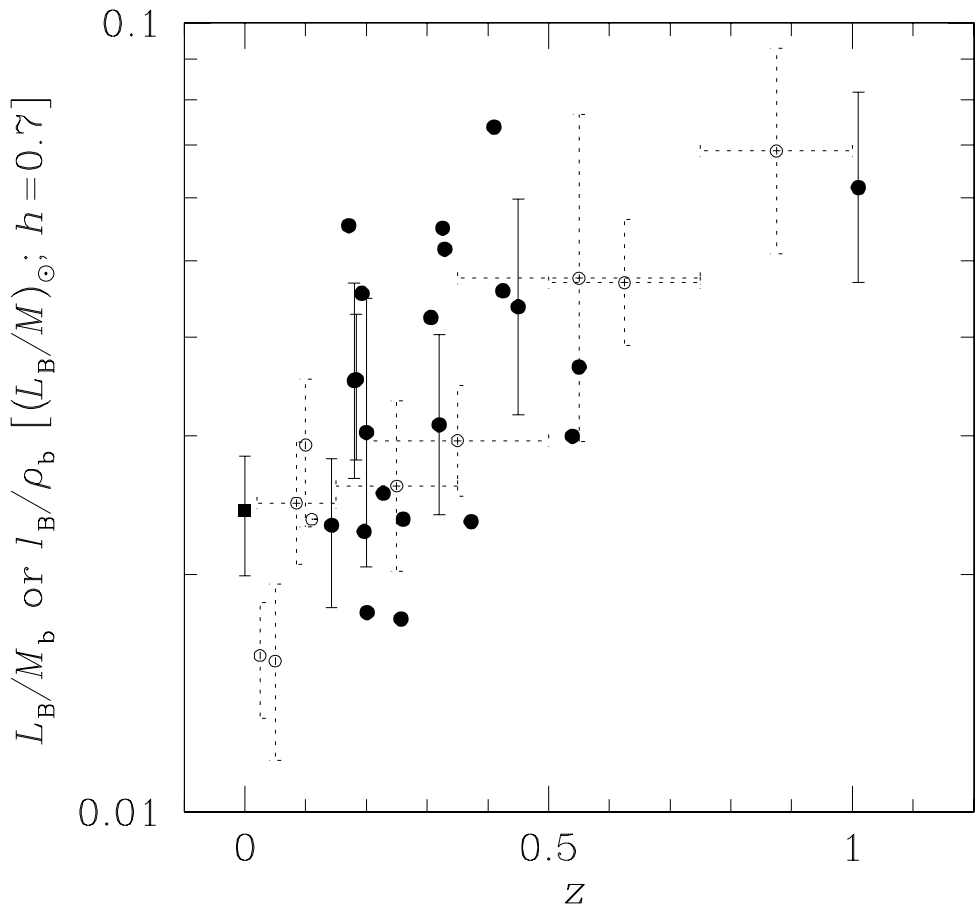
\end{document}